\documentclass[10pt,pra,twocolumn,floatfix]{revtex4}

\usepackage{amsfonts}
\usepackage{amssymb}
\usepackage{amsmath}
\usepackage[dvips]{graphicx}

\begin{document}

\title{Gate sequence for continuous variable one-way quantum computation}
\author{Xiaolong Su, Shuhong Hao, Xiaowei Deng, Lingyu Ma, Meihong Wang,
Xiaojun Jia, Changde Xie and Kunchi Peng}
\email{kcpeng@sxu.edu.cn}
\affiliation{State Key Laboratory of Quantum Optics and Quantum Optics Devices,\\
Institute of Opto-Electronics, Shanxi University, Taiyuan, 030006, People's
Republic of China}
\maketitle

\textbf{Measurement-based one-way quantum computation (QC) using cluster
states as resources provides an efficient model to perform computation and
information processing of quantum codes. Arbitrary Gaussian QC can be
implemented by sufficiently long single-mode and two-mode gate sequences.
However, continuous variable (CV) gate sequences have not been realized so
far due to an absence of cluster states larger than four submodes. Here we
present the first CV gate sequence consisting of a single-mode squeezing
gate and a two-mode controlled-phase gate based on a six-mode cluster state.
The quantum property of this gate sequence is confirmed by the fidelities
and the quantum entanglement of two output modes, which depend on both the
squeezing and controlled-phase gates. The experiment demonstrates the
feasibility of implementing Gaussian QC by means of accessible gate
sequences.}

Measurement-based one-way QC performs computation by measurement and
classical feedforward on a multipartite cluster entangled state \cite%
{Raussendorf2001}. One-way QC was first demonstrated using four-photon
cluster states \cite{Walther2005,Chen}. Besides photonic systems \cite%
{Walther2005,Chen,Brien,Brien2009,Tok}, other QC systems with discrete
quantum variables, such as ionic \cite{Cirac,Mon,Kie,Kim}, superconducting
\cite{Mak,Yam,Nee,Di,Bi}, and solid state \cite{Ber,Fus,Han} systems, have
been investigated.

In parallel, the theoretical and experimental explorations on one-way CVQC
have also been developed \cite%
{Menicucci2006,Weedbrook2012,Zhang2006,Loock2007,vanLoockJOSA2007,Gu2009,Zwi,Fer,Miwa2009,Wang2010,Ukai2011,Ukai20112}%
. In contrast to the probabilistic generation of photonic qubits in most
cases, CV cluster states are produced in an unconditional fashion and thus
the one-way QC with CV cluster states can be implemented deterministically
\cite%
{Su2007,Yukawa2008,Tan2008,Matt2011,Ukai2011,Wang2010,Ukai20112,Miwa2009}.
Although individual single-mode and two-mode logical gates towards
implementing multimode Gaussian transformation in a one-way CVQC fashion
have been experimentally demonstrated by using four-mode cluster entangled
states of light \cite{Miwa2009,Wang2010,Ukai2011,Ukai20112}, CV gate
sequences consisting of different single logical elements, which are
necessary for realizing practical QC, have not been reported up to now. It
is now important to investigate the gate sequences for QC as sufficiently
large cluster states have recently been prepared, including eight-photon
\cite{Huang2011,Yao2012}, eight-quantum mode \cite{Su2012,Arm2012} and up to
10,000-qumode \cite{Yok2013} optical cluster states.

Here, we design and experimentally accomplish a CV gate sequence, in which a
single-mode squeezing gate and a two-mode controlled-phase (CZ) gate are
cascaded. A vacuum optical input signal is first squeezed by the squeezing
gate and successively the squeezed output enters the CZ gate as one of its
two inputs. A vacuum state or a squeezed state of light produced by an
off-line nondegenerate optical parametric amplifier (NOPA) is used for the
other CZ gate input. The experimental result shows that after two
independent input states are transformed by the gate sequence, the two
output states produced are entangled and their fidelities are better than
that obtained by using coherent states as resources. Our experiments also
prove that the entanglement degree and the fidelity depend simultaneously on
two cascaded logical gates, which manifests the sequence feature of the
presented system. Besides the gate sequences only involving multimode
Gaussian transformation, for demonstrating universal one-way CVQC, at least
a non-Gaussian operation is required \cite{Menicucci2006}. Many theoretical
protocols and schemes on the universal CVQC have been proposed \cite%
{Menicucci2006,Weedbrook2012}.

\section*{Results}

\subsection*{Preparation of six-mode CV cluster states}

CV cluster states are defined as \cite{Zhang2006,Loock2007}
\begin{equation}
\hat{p}_{a}-\sum_{b\in N_{a}}\hat{x}_{b}\equiv \hat{\delta}_{a}\rightarrow
0,\ \ a\in G.
\end{equation}%
In the limit of infinite squeezing, the $N$-mode cluster states are a
simultaneous zero eigenstate of the $N$\ linear combinations in Eq. (1).
Here the amplitude ($\hat{x}$) and phase ($\hat{p}$) quadratures of an
optical mode $\hat{a}$\ are defined as $\hat{x}=(\hat{a}+\hat{a}^{\dagger
})/2$\ and $\hat{p}=(\hat{a}-\hat{a}^{\dagger })/2i$. The modes $a\in G$\
denote the vertices of the graph $G$, while the modes $b\in N_{a}$\ are the
nearest neighbors of mode $\hat{a}$. One time measurement on cluster state
can not destroy the entanglement totally, which means that cluster state has
a strong property of entanglement persistence.

A general way to build CV cluster state is to implement an appropriate
unitary transformation ($U)$\ on a series of input $\hat{p}$-squeezed
states, $\hat{a}_{l}=e^{+r}\hat{x}_{l}^{(0)}+ie^{-r}\hat{p}_{l}^{(0)}$,
where $r$\ is the squeezing parameter, $\hat{x}^{(0)}$\ and $\hat{p}^{(0)}$\
represent the quadratures of a vacuum state, which has a noise variance $%
\left\langle \Delta ^{2}\hat{x}^{(0)}\right\rangle =\left\langle \Delta ^{2}%
\hat{p}^{(0)}\right\rangle =1/4$. According to a general linear optics
transformation $\hat{b}_{k}=\sum\nolimits_{l}U_{kl}\hat{a}_{l}$, the output
modes can be obtained \cite{Loock2007}. The transformation matrix $U$\ can
be decomposed into a network of beam-splitters, which corresponds to the
experimental system for generating required CV cluster state. We designed
the beam-splitter network of producing CV six-mode linear cluster state with
three NOPAs, as shown in Fig. 1. A NOPA can simultaneously generate a $\hat{x%
}$-squeezed state and a $\hat{p}$-squeezed state \cite{Yun2000}. The three $%
\hat{x}$-squeezed states and three $\hat{p}$-squeezed states prepared by the
three NOPAs, are denoted by $\hat{a}_{1},\hat{a}_{3},\hat{a}_{5}$, $\hat{a}%
_{m}=e^{-r}\hat{x}_{m}^{(0)}+ie^{+r}\hat{p}_{m}^{(0)},(m=1,3,5)$\ for $\hat{x%
}$-squeezed states, and $\hat{a}_{2},\hat{a}_{4},\hat{a}_{6}$, $\hat{a}%
_{n}=e^{+r}\hat{x}_{n}^{(0)}+ie^{-r}\hat{p}_{n}^{(0)},(n=2,4,6)$\ for $\hat{p%
}$-squeezed states, respectively. Here we have assumed that all squeezed
states produced by the three NOPAs have identical squeezing degree due to
the equality of their configuration (see Methods). We deduce the
transformation matrix for generating CV six-mode linear cluster state using
three $\hat{x}$-squeezed states and three $\hat{p}$-squeezed states as input
states, which is given by \cite{Loock2007}

\begin{equation}
U_{6}=\left(
\begin{array}{cccccc}
\frac{i}{\sqrt{2}} & \frac{i}{\sqrt{3}} & -\sqrt{\frac{2}{39}} & -\sqrt{%
\frac{3}{26}} & 0 & 0 \\
-\frac{1}{\sqrt{2}} & \frac{1}{\sqrt{3}} & i\sqrt{\frac{2}{39}} & i\sqrt{%
\frac{3}{26}} & 0 & 0 \\
0 & \frac{i}{\sqrt{3}} & 2\sqrt{\frac{2}{39}} & \sqrt{\frac{6}{13}} & 0 & 0
\\
0 & 0 & -i\sqrt{\frac{6}{13}} & 2i\sqrt{\frac{2}{39}} & -\frac{1}{\sqrt{3}}
& 0 \\
0 & 0 & \sqrt{\frac{3}{26}} & -\sqrt{\frac{2}{39}} & \frac{i}{\sqrt{3}} & -%
\frac{i}{\sqrt{2}} \\
0 & 0 & i\sqrt{\frac{3}{26}} & -i\sqrt{\frac{2}{39}} & -\frac{1}{\sqrt{3}} &
-\frac{1}{\sqrt{2}}%
\end{array}%
\right) .
\end{equation}

The matrix can be decomposed into $%
U_{6}=F_{1}I_{2}(-1)F_{3}F_{4}F_{5}I_{6}(-1)B_{56}^{+}(1/2)\allowbreak
F_{5}B_{12}^{+}(1/2)B_{45}^{+}(2/3)F_{5}B_{23}^{+}(2/3)F_{3}B_{34}^{-}(4/13)$%
, where $F_{k}$\ denotes the Fourier transformation of mode $k$, which
corresponds to a 90$^{\circ }$\ rotation in the phase space; $B_{kl}^{\pm
}(T_{j})$\ stands for the linearly optical transformation on the jth
beam-splitter with the transmission of $T_{j}$\ ($j=1,\ldots 5$), where $%
(B_{kl}^{\pm })_{kk}=\sqrt{1-T}$, $(B_{kl}^{\pm })_{kl}=\sqrt{T}$, $%
(B_{kl}^{\pm })_{lk}=\pm \sqrt{T}$, and $(B_{kl}^{\pm })_{ll}=\mp \sqrt{1-T}%
, $\ are elements of beam-splitter matrix. $I_{k}(-1)=e^{i\pi }$\
corresponds to a 180$^{\circ }$\ rotation of mode $k$\ in phase space. When
the transmissions of the five beam-splitters are chosen as $T_{1}=4/13$, $%
T_{2}=T_{3}=2/3$, $T_{4}=T_{5}=1/2$, the six output optical modes construct
a six-mode CV linear cluster state. The corresponding excess noise terms for
each mode of the CV six-mode linear cluster state are respectively denoted
by
\begin{eqnarray}
\hat{\delta}_{1} &=&\sqrt{2}e^{-r}\hat{x}_{1}^{(0)},  \notag \\
\hat{\delta}_{2} &=&\sqrt{3}e^{-r}\hat{p}_{2}^{(0)},  \notag \\
\hat{\delta}_{3} &=&\frac{1}{\sqrt{2}}e^{-r}\hat{x}_{1}^{(0)}+\sqrt{\frac{13%
}{6}}e^{-r}\hat{p}_{4}^{(0)}+\frac{1}{\sqrt{3}}e^{-r}\hat{x}_{5}^{(0)},
\notag \\
\hat{\delta}_{4} &=&\frac{1}{\sqrt{3}}e^{-r}\hat{p}_{2}^{(0)}-\sqrt{\frac{13%
}{6}}e^{-r}\hat{x}_{3}^{(0)}+\frac{1}{\sqrt{2}}e^{-r}\hat{p}_{6}^{(0)},
\notag \\
\hat{\delta}_{5} &=&\sqrt{3}e^{-r}\hat{x}_{5}^{(0)},  \notag \\
\hat{\delta}_{6} &=&\sqrt{2}e^{-r}\hat{p}_{6}^{(0)}.
\end{eqnarray}%
Obviously, in the ideal case with infinite squeezing ($r\rightarrow \infty $%
), these excess noises will vanish and the better the squeezing, the smaller
the noise term.

\begin{figure}[tbp]
\centerline{
\includegraphics[width=85mm]{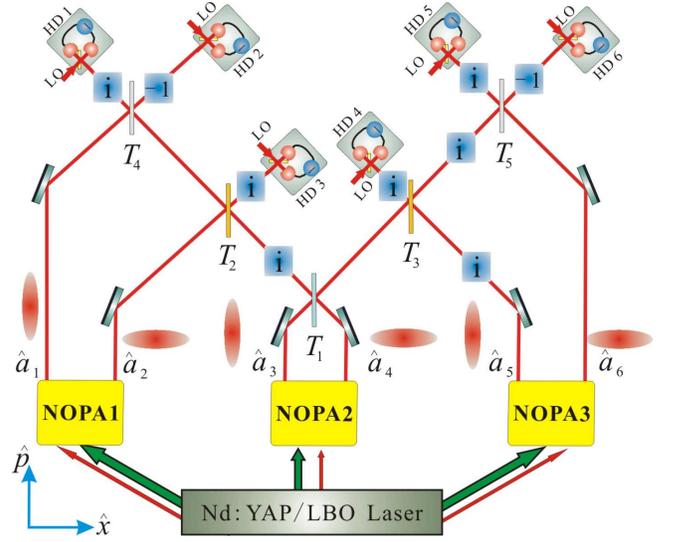}
}
\caption{\textbf{The schematic of six-mode CV cluster state generation
system. }$T$: transmission efficient of beam splitter, Boxes including $i$
are Fourier transforms ($90^{\circ }$ rotations in phase space), and $-1$ is
a $180^{\circ }$ rotation, LO: local osillation beam, HD: homodyne detector.}
\end{figure}

\begin{figure}[tbp]
\centerline{
\includegraphics[width=85mm]{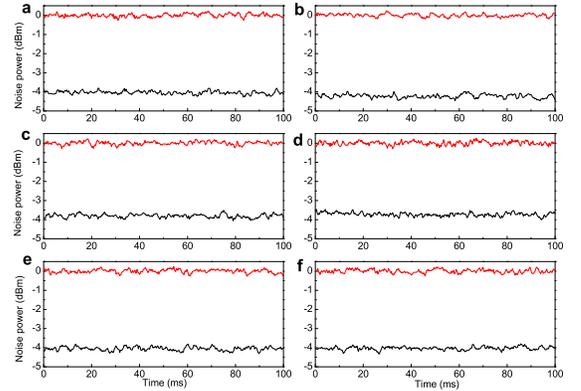}
}
\caption{\textbf{The measured correlation noises of six-mode CV cluster
state.} The red and black lines in all graphs are shot noise level and
correlation variances of nullifiers, respectively. (a)-(f) are noise powers
of $\left\langle \Delta ^{2}(\hat{p}_{1}-\hat{x}_{2})\right\rangle $, $%
\left\langle \Delta ^{2}(\hat{p}_{2}-\hat{x}_{1}-\hat{x}_{3})\right\rangle $%
, $\left\langle \Delta ^{2}(\hat{p}_{3}-\hat{x}_{2}-\hat{x}%
_{4})\right\rangle $, $\left\langle \Delta ^{2}(\hat{p}_{4}-\hat{x}_{3}-\hat{%
x}_{5})\right\rangle $, $\left\langle \Delta ^{2}(\hat{p}_{5}-\hat{x}_{4}-%
\hat{x}_{6})\right\rangle $, and $\left\langle \Delta ^{2}(\hat{p}_{6}-\hat{x%
}_{5})\right\rangle $, respectively. Measurement frequency is 2 MHz, the
spectrum analyzer resolution bandwidth is 30 kHz, and the video bandwidth is
100 Hz.}
\end{figure}

Fig. 2 shows the experimental results of six-mode CV cluster state. Red and
black lines correspond to shot-noise-level (SNL) and quantum correlation
noise, respectively. The measured noise are $\left\langle \Delta ^{2}(\hat{p}%
_{1}-\hat{x}_{2})\right\rangle =-4.04\pm 0.09$\ dB, $\left\langle \Delta
^{2}(\hat{p}_{2}-\hat{x}_{1}-\hat{x}_{3})\right\rangle =-4.22\pm 0.10$\ dB, $%
\left\langle \Delta ^{2}(\hat{p}_{3}-\hat{x}_{2}-\hat{x}_{4})\right\rangle
=-3.80\pm 0.10$\ dB, $\left\langle \Delta ^{2}(\hat{p}_{4}-\hat{x}_{3}-\hat{x%
}_{5})\right\rangle =-3.72\pm 0.09$\ dB, $\left\langle \Delta ^{2}(\hat{p}%
_{5}-\hat{x}_{4}-\hat{x}_{6})\right\rangle =-4.05\pm 0.10$\ dB, and $%
\left\langle \Delta ^{2}(\hat{p}_{6}-\hat{x}_{5})\right\rangle =-4.03\pm
0.09 $\ dB, respectively.

According to the inseparability criteria for CV multipartite entangled
states proposed by van Loock and Furusawa \cite{Loock2003}, we deduced the
inseparability criteria for CV six-partite linear cluster state, which are
\begin{eqnarray}
\left\langle \Delta ^{2}(\hat{p}_{1}-\hat{x}_{2})\right\rangle +\left\langle
\Delta ^{2}(\hat{p}_{2}-\hat{x}_{1}-\hat{x}_{3})\right\rangle &<&1, \\
\left\langle \Delta ^{2}(\hat{p}_{2}-\hat{x}_{1}-\hat{x}_{3})\right\rangle
+\left\langle \Delta ^{2}(\hat{p}_{3}-\hat{x}_{2}-\hat{x}_{4})\right\rangle
&<&1, \\
\left\langle \Delta ^{2}(\hat{p}_{3}-\hat{x}_{2}-\hat{x}_{4})\right\rangle
+\left\langle \Delta ^{2}(\hat{p}_{4}-\hat{x}_{3}-\hat{x}_{5})\right\rangle
&<&1, \\
\left\langle \Delta ^{2}(\hat{p}_{4}-\hat{x}_{3}-\hat{x}_{5})\right\rangle
+\left\langle \Delta ^{2}(\hat{p}_{5}-\hat{x}_{4}-\hat{x}_{6})\right\rangle
&<&1, \\
\left\langle \Delta ^{2}(\hat{p}_{5}-\hat{x}_{4}-\hat{x}_{6})\right\rangle
+\left\langle \Delta ^{2}(\hat{p}_{6}-\hat{x}_{5})\right\rangle &<&1.
\end{eqnarray}

Substituting the measured quantum noise (Fig. 2) into Eqs. (4)-(8), we can
calculate the combinations of the correlation variances, which are 0.48,
0.59, 0.63, 0.62 and 0.50, respectively. Since all these values are smaller
than the boundary of 1, the prepared six quantum modes satisfy the
inseparability criteria and form an six-mode cluster entangled state.

\subsection*{Configuration of the gate sequence}

\begin{figure}[tbp]
\setlength{\belowcaptionskip}{-3pt}
\centerline{
\includegraphics[width=\columnwidth]{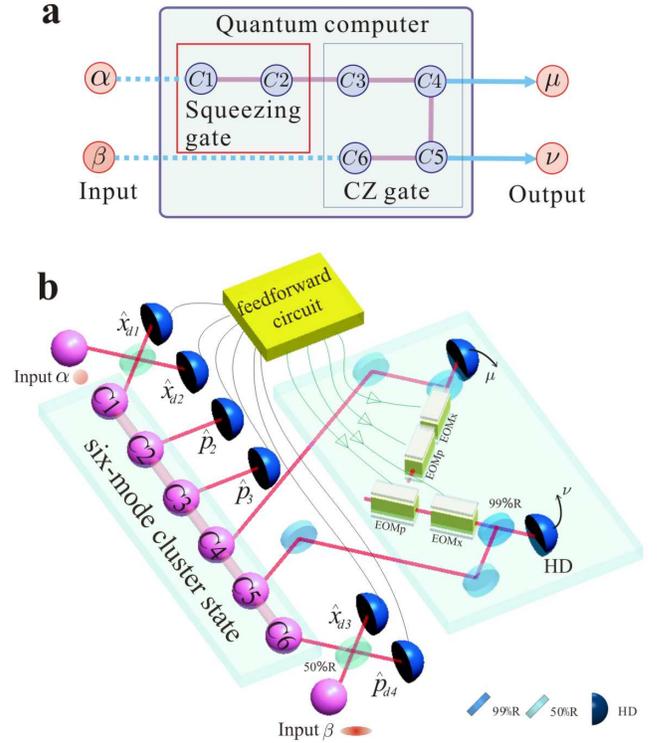}
}
\caption{\textbf{Schematic of the gate sequence.} \textbf{a}, The graph
representation. The input state $\protect\alpha $ is coupled to a six-mode
CV cluster state $C1$-$C2$-$C3$-$C4$-$C5$-$C6$. Squeezing operation is
performed on input mode $\protect\alpha $, then CZ gate is performed on the
input state $\protect\beta $ and output of the squeezing operation. \textbf{b%
}, schematic of the experimental setup. The input states $\protect\alpha $
and $\protect\beta $ are coupled to a six-mode CV cluster state via $50\%$
beam-splitter (50\%R). Measurement results from homodyne detection (HD)
systems are feedforward to modes $C4$ and $C5$. The output modes $\protect%
\mu $ and $\protect\nu $ are measured by two HD systems. EOMx and EOMp:
amplitude and phase electro-optical modulators. 99\%R: a mirror with 99\%
reflection coefficient.}
\end{figure}

As shown in Fig. 3a, we demonstrate the gate sequence of a squeezing gate
and a CZ gate using a six-mode cluster state as ancillary state. First, we
perform squeezing gate on input mode $\alpha $ (target mode) to produce a
phase squeezed state. Then a CZ gate is performed on the output mode from
the squeezing gate and the other input mode $\beta $ (control mode) coming
from outside of the sequence.

The single-mode squeezing gate is expressed by $\hat{S}(r_{s})=e^{ir_{s}(%
\hat{x}\hat{p}+\hat{p}\hat{x})}$. The input-output relation of the
single-mode squeezing gate is $\mathbf{\hat{\xi}}_{j}^{\prime }=\mathbf{S%
\hat{\xi}}_{j}$, where $\mathbf{\hat{\xi}}_{j}=$ $(\hat{x}_{j},\hat{p}%
_{j})^{T}$ and
\begin{equation}
\mathbf{S=}%
\begin{pmatrix}
e^{r_{s}} & 0 \\
0 & e^{-r_{s}}%
\end{pmatrix}%
\end{equation}%
represents the squeezing operation on phase quadrature. The CV CZ gate
corresponds to the unitary operator $\hat{C}_{Zjk}=e^{2i\hat{x}_{j}\hat{x}%
_{k}}$ with the input-output relation,
\begin{equation}
\mathbf{\hat{\xi}}_{jk}^{\prime }=%
\begin{pmatrix}
\mathbf{I} & \mathbf{C} \\
\mathbf{C} & \mathbf{I}%
\end{pmatrix}%
\mathbf{\hat{\xi}}_{jk},
\end{equation}%
where $\mathbf{\hat{\xi}}_{jk}=$ $(\hat{x}_{j},\hat{p}_{j},\hat{x}_{k},\hat{p%
}_{k})^{T}$,
\begin{equation}
\mathbf{C}=%
\begin{pmatrix}
0 & 0 \\
1 & 0%
\end{pmatrix}%
,
\end{equation}%
and $\mathbf{I}$ is a $2\times 2$ identity matrix.

The transformation matrix of the gate sequence is given by
\begin{equation}
\mathbf{U}=%
\begin{pmatrix}
\mathbf{I} & \mathbf{C} \\
\mathbf{C} & \mathbf{I}%
\end{pmatrix}%
\text{\textperiodcentered }%
\begin{pmatrix}
\mathbf{S} & 0 \\
0 & \mathbf{I}%
\end{pmatrix}%
.
\end{equation}%
The input-output relation of the gate sequence is expressed by $\mathbf{\hat{%
\xi}}_{\mu \nu }=\mathbf{U\hat{\xi}}_{\alpha \beta }$ in the ideal case.
However, in any practical cases, the excess noise coming from imperfect
entanglement of the CV cluster state will inevitably affect the performances
of the logical gates, thus a noise term $\mathbf{\hat{\delta}}$ should be
added, that is
\begin{equation}
\mathbf{\hat{\xi}}_{\mu \nu }=\mathbf{U\hat{\xi}}_{\alpha \beta }+\mathbf{%
\hat{\delta}},
\end{equation}%
where $\mathbf{\hat{\delta}=(}\hat{\delta}_{1}-\hat{\delta}_{3},\hat{\delta}%
_{4}-\hat{\delta}_{2}-\hat{\delta}_{6},-\hat{\delta}_{6},\hat{\delta}_{1}+%
\hat{\delta}_{5}-\hat{\delta}_{3}\mathbf{)}$ represents all excess noises of
the gate sequence.

The schematic of experimental setup is shown in Fig. 3b. A six-mode cluster
state involving six submodes $C1$, $C2$, $C3$, $C4$, $C5$, and $C6$ with the
squeezing about $-$4.0 dB on average is used as the resource (ancillary)
state. The input states $\alpha $ and $\beta $ are coupled to submodes $C1$
and $C6$ by two 50\% beam-splitters respectively. The measurement results of
the output modes from the two 50\% beam-splitters as well as the submodes $C2
$ and $C3$ are feedforward to submodes $C4$ and $C5$ through classical
feedforward circuits. In the operation of gate sequence, the measured
observables are $\hat{x}_{d1}=[\cos \theta _{1}(\hat{x}_{\alpha }-\hat{x}%
_{1})+\sin \theta _{1}(\hat{p}_{\alpha }-\hat{p}_{1})]/\sqrt{2}$, $\hat{x}%
_{d2}=[\cos \theta _{2}(\hat{x}_{\alpha }+\hat{x}_{1})+\sin \theta _{2}(\hat{%
p}_{\alpha }+\hat{p}_{1})]/\sqrt{2}$, $\hat{p}_{2}$, $\hat{p}_{3}$, $\hat{x}%
_{d3}=(\hat{x}_{\beta }-\hat{p}_{6})/\sqrt{2}$, and $\hat{p}_{d4}=(\hat{p}%
_{\beta }-\hat{x}_{6})/\sqrt{2}$. The measurement angle $\theta _{1}$ and $%
\theta _{2}$ in the homodyne detection for $\hat{x}_{d1}$ and $\hat{x}_{d2}$
determine the squeezing operation according to the transformation matrix
\begin{equation}
\mathbf{S=}%
\begin{pmatrix}
\cot \theta _{2} & 0 \\
0 & \tan \theta _{2}%
\end{pmatrix}%
,
\end{equation}%
if we choose $\theta _{2}=-\theta _{1}$. $\mathbf{S}$ corresponds to a
single-mode amplitude and phase squeezing gate when $\cot \theta
_{2}=e^{-r_{s}}$ and $e^{r_{s}}$, respectively. The measurement angles ($%
\theta _{1}$, $\theta _{2}$) for 0 dB, $-$3 dB, $-$6 dB, $-$9 dB and $-$12
dB squeezing are ($-45.00%
{{}^\circ}%
$, $45.00%
{{}^\circ}%
$), ($-35.30%
{{}^\circ}%
$, $35.30%
{{}^\circ}%
$), ($-26.62%
{{}^\circ}%
$, $26.62%
{{}^\circ}%
$), ($-19.54%
{{}^\circ}%
$, $19.54%
{{}^\circ}%
$), and ($-14.10%
{{}^\circ}%
$, $14.10%
{{}^\circ}%
$), respectively. These measurement results are feedforwarded to the
amplitude and phase quadratures of modes $C4$ and $C5$ via electro-optical
modulators (EOM), respectively. The corresponding feedforward operation is
expressed by $\hat{X}_{C4}(f1)\hat{Z}_{C4}(f2)\hat{X}_{C5}(f3)\hat{Z}%
_{C5}(f4)$, where $\hat{X}_{j}(s)=e^{-2is\hat{p}_{j}}$ and $\hat{Z}%
_{j}(s)=e^{2is\hat{x}_{j}}$ are the amplitude and phase displacement
operators, respectively,\ $f1=-\hat{p}_{3}+\frac{\csc \theta _{2}}{\sqrt{2}}%
\hat{x}_{d1}+\frac{\csc \theta _{2}}{\sqrt{2}}\hat{x}_{d2}$, $f2=-\hat{p}%
_{2}+\sqrt{2}\hat{x}_{d3}-\frac{\sec \theta _{2}}{\sqrt{2}}\hat{x}_{d1}+%
\frac{\sec \theta _{2}}{\sqrt{2}}\hat{x}_{d2}$, $f3=\sqrt{2}\hat{x}_{d3}$,
and $f4=-\hat{p}_{3}+\sqrt{2}\hat{p}_{d4}+\frac{\csc \theta _{2}}{\sqrt{2}}%
\hat{x}_{d1}+\frac{\csc \theta _{2}}{\sqrt{2}}\hat{x}_{d2}$. The output
modes are detected by two homodyne detection systems. The quadrature
components of the output modes for the gate sequence are given by{\small
\begin{gather}
\begin{pmatrix}
\hat{x}_{\mu } \\
\hat{p}_{\mu } \\
\hat{x}_{\nu } \\
\hat{p}_{\nu }%
\end{pmatrix}%
=%
\begin{pmatrix}
\hat{x}_{C4} \\
\hat{p}_{C4} \\
\hat{x}_{C5} \\
\hat{p}_{C5}%
\end{pmatrix}%
+G\left(
\begin{pmatrix}
\hat{p}_{3} \\
\hat{p}_{2} \\
\sqrt{2}\hat{x}_{d3} \\
\sqrt{2}\hat{p}_{d4}%
\end{pmatrix}%
-%
\begin{pmatrix}
\frac{\csc \theta _{2}}{\sqrt{2}} & \frac{\csc \theta _{2}}{\sqrt{2}} \\
\frac{-\sec \theta _{2}}{\sqrt{2}} & \frac{\sec \theta _{2}}{\sqrt{2}} \\
0 & 0 \\
0 & 0%
\end{pmatrix}%
\dbinom{\hat{x}_{d1}}{\hat{x}_{d2}}\right)   \notag \\
=%
\begin{pmatrix}
1 & 0 & 0 & 0 \\
0 & 1 & 1 & 0 \\
0 & 0 & 1 & 0 \\
1 & 0 & 0 & 1%
\end{pmatrix}%
\begin{pmatrix}
\cot \theta _{2} & 0 & 0 & 0 \\
0 & \tan \theta _{2} & 0 & 0 \\
0 & 0 & 1 & 0 \\
0 & 0 & 0 & 1%
\end{pmatrix}%
\begin{pmatrix}
\hat{x}_{\alpha } \\
\hat{p}_{\alpha } \\
\hat{x}_{\beta } \\
\hat{p}_{\beta }%
\end{pmatrix}%
+%
\begin{pmatrix}
\hat{\delta}_{1}-\hat{\delta}_{3} \\
-\hat{\delta}_{6}+\hat{\delta}_{4}-\hat{\delta}_{2} \\
-\hat{\delta}_{6} \\
\hat{\delta}_{5}-\hat{\delta}_{3}+\hat{\delta}_{1}%
\end{pmatrix}%
,  \notag \\
\end{gather}%
} where%
\begin{equation}
G=%
\begin{pmatrix}
-1 & 0 & 0 & 0 \\
0 & -1 & 1 & 0 \\
0 & 0 & 1 & 0 \\
-1 & 0 & 0 & 1%
\end{pmatrix}%
\end{equation}%
is the feedforward gain factor. From Eqs. (14) and (15) we obtain
\begin{eqnarray}
\hat{x}_{\mu } &=&\hat{x}_{\alpha }e^{r_{s}}+\hat{\delta}_{1}-\hat{\delta}%
_{3}, \\
\hat{p}_{\mu } &=&\hat{p}_{\alpha }e^{-r_{s}}+\hat{x}_{\beta }-\hat{\delta}%
_{2}+\hat{\delta}_{4}-\hat{\delta}_{6},  \notag \\
\hat{x}_{\nu } &=&\hat{x}_{\beta }-\hat{\delta}_{6},  \notag \\
\hat{p}_{\nu } &=&\hat{p}_{\beta }+\hat{x}_{\alpha }e^{r_{s}}+\hat{\delta}%
_{1}+\hat{\delta}_{5}-\hat{\delta}_{3}.  \notag
\end{eqnarray}%
After a vacuum signal ($\alpha $) passes through the squeezing gate, its
phase ($\hat{p}_{\alpha }$) and amplitude ($\hat{x}_{\alpha }$) are squeezed
($\hat{p}_{\alpha }e^{-r_{s}}$) and anti-squeezed ($\hat{x}_{\alpha
}e^{r_{s}}$), respectively. Then, the squeezed signal passes through the CZ
gate, as the usual result of CZ gate \cite{Ukai20112}, the anti-squeezed
amplitude signal is transformed into the phase quadrature ($\hat{p}_{\nu }$%
)\ of the resultant output mode $\nu $ and its amplitude quadrature ($\hat{x}%
_{\nu }$)\ keeps unchanging in the ideal case without excess noises ($\hat{%
\delta}_{1-6}=0$).

\begin{figure}[tbp]
\setlength{\belowcaptionskip}{-3pt}
\centerline{
\includegraphics[width=\columnwidth]{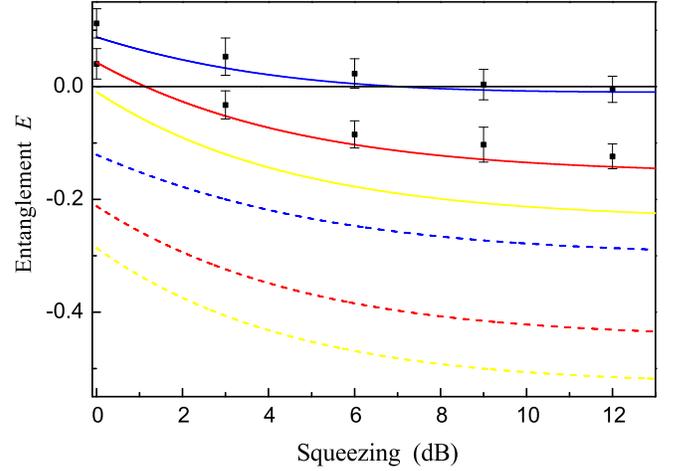}
}
\caption{\textbf{Dependence of entanglement on squeezing of the squeezing
gate.} The dependence of the entanglement between the gate sequence output
modes on the squeezing of the squeezing gate is plotted here. The blue, red
and yellow lines correspond to the input mode $\protect\beta $ being a
vacuum state, $-$4 dB and $-$12 dB phase squeezed state, respectively. The
solid and dashed lines correspond to $-$4 dB and $-$6 dB squeezing of the
six-mode cluster state, respectively. The black dots represent the
experimental data. Error bars represent $\pm$ one standard deviation and are
obtained based on the statistics of the measured noise variances.}
\end{figure}

\subsection*{Experimental results}

Since CZ gate is a two-mode entangling gate, its quantum character can be
verified by the sufficient condition of inseparability for a two-mode state
\cite{Duan}, that is
\begin{equation}
\left\langle \Delta ^{2}(g\hat{p}_{\mu }-\hat{x}_{\nu })\right\rangle
+\left\langle \Delta ^{2}(g\hat{p}_{\nu }-\hat{x}_{\mu })\right\rangle <g,
\end{equation}%
where $g$ is the optimal gain factor which makes the left side of the
inequality minimum. By calculating the minimal value of equation (16), the
optimal gain is obtained
\begin{equation}
g=\frac{e^{2r_{\beta }}(3+2e^{2r}+e^{2r+2r_{\beta }}+e^{2r}\cot ^{2}\theta
_{2})}{e^{2r}+8e^{2r_{\beta }}+e^{2r+4r_{\beta }}+e^{2r+2r_{\beta }}(\cot
^{2}\theta _{2}+\tan ^{2}\theta _{2})},
\end{equation}%
where $r_{\beta }$\ is the squeezing parameter of the input mode $\beta $.
Fig. 4 shows the dependence of entanglement degree between the two output
states of the gate sequence on squeezing degree of the squeezing gate with
an input vacuum mode $\alpha $ for three different $\beta $ states (blue:
vacuum state, red: $-4$ dB phase squeezed state, yellow: $-12$ dB phase
squeezed state). The entanglement degree is quantified by $E=\left\langle
\Delta ^{2}(g\hat{p}_{\mu }-\hat{x}_{\nu })\right\rangle +\left\langle
\Delta ^{2}(g\hat{p}_{\nu }-\hat{x}_{\mu })\right\rangle -g$. When $E<0$ ,
the entanglement exists and the smaller the $E$, the stronger the
entanglement. The solid and dashed lines correspond to $-4$ and $-6$ dB
squeezing of the six-mode cluster state, respectively. All traces are
plotted according to the real loss in our experimental system (see Methods).
We can see that the entanglement degree between the output states ($\mu $
and $\nu $) not only depends on the operation of CZ gate, but also is
controlled by the squeezing operation of the squeezing gate. For a given $%
\beta $ state, when the squeezing of the squeezing gate increases the
entanglement degree of output states increases. On the other hand, the
operation of CZ gate also depends on the feature of $\beta $ state. The
phase squeezing of $\beta $ state will improve the entanglement degree of
the output modes. Of course, the largest influence to the capacity of the
gate sequence comes from the quality of the resource state. When the
squeezing of the six-mode cluster state increases from $-4$ dB (solid lines)
to $-6$ dB (dashed lines) the entanglement degree of output states will be
significantly improved. Since $-4$ dB cluster squeezing is available in our
experiment, the obtained maximal entanglement degree is only $-0.005$ for
the case of using two vacuum states to be $\alpha $ and $\beta $ (solid blue
line). If $\beta $ is a phase squeezed state (red and yellow lines) the
entanglement will increase for the same squeezing degree of the squeezing
gate and an identical resource state. The experimentally measured data
points obtained under different measurement angles of the squeezing gate are
marked by little black dots with error bars, which shows that the
experimental results are in reasonable agreement with the theoretical
expectation. The experimentally measured entanglement degrees, fidelities,
and corresponding optimal gain factors are listed in table 1. The
entanglement degree of output modes depends on both operations of two
cascaded gates, which clearly shows that the final results are decided by
the gate sequence.

\begin{table*}[tbp]
\begin{tabular}{|lllll|}
\hline
\multicolumn{5}{|l|}{\textbf{Table 1} $\shortmid $ \textbf{The entanglement
and fidelity at experimentally measured data points.}} \\ \hline
\textbf{Data point}\ \ \ \  & $\mathbf{g\qquad }$ & $\mathbf{E}$ & $\mathbf{F%
}_{\mu }$ & $\mathbf{F}_{\nu }$ \\ \hline
a & $0.72\qquad \qquad $ & $0.112\pm 0.026\qquad \qquad $ & $0.832\pm
0.011\qquad \qquad $ & $0.873\pm 0.013\qquad $ \\
b & $0.81$ & $0.053\pm 0.033$ & $0.882\pm 0.011$ & $0.902\pm 0.014$ \\
c & $0.87$ & $0.023\pm 0.026$ & $0.905\pm 0.009$ & $0.942\pm 0.012$ \\
d & $0.92$ & $0.004\pm 0.027$ & $0.888\pm 0.009$ & $0.951\pm 0.011$ \\
e & $0.95$ & $-0.005\pm 0.024$ & $0.886\pm 0.012$ & $0.956\pm 0.009$ \\
f & $0.83$ & $0.040\pm 0.026$ & $0.860\pm 0.013$ & $0.854\pm 0.013$ \\
g & $0.90$ & $-0.033\pm 0.029$ & $0.903\pm 0.014$ & $0.891\pm 0.013$ \\
h & $0.94$ & $-0.085\pm 0.024$ & $0.922\pm 0.009$ & $0.934\pm 0.009$ \\
i & $0.96$ & $-0.103\pm 0.031$ & $0.932\pm 0.011$ & $0.950\pm 0.010$ \\
j & $0.98$ & $-0.124\pm 0.022$ & $0.923\pm 0.006$ & $0.947\pm 0.006$ \\
\hline
\multicolumn{5}{|l|}{a-e: $\alpha $ and $\beta $ are vacuum state, squeezing
of the squeezing gate are 0, $-3$,$-6$, $-9$ and $-12$ dB,} \\
\multicolumn{5}{|l|}{respectively. f-j: $\alpha $ is a vacuum state, $\beta $
is a $-4$ dB phase squeezed state, squeezing of the} \\
\multicolumn{5}{|l|}{squeezing gate are 0, $-3$, $-6$, $-9$ and $-12$ dB,
respectively.} \\ \hline
\end{tabular}%
\end{table*}

\begin{figure}[tbp]
\par
\begin{center}
\includegraphics[width=\columnwidth]{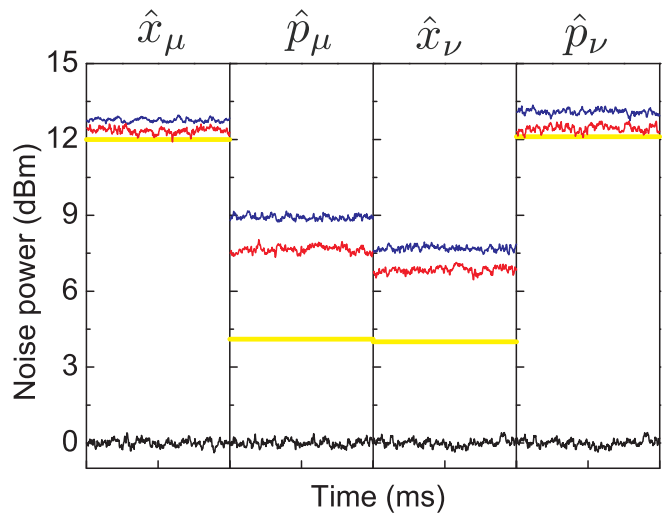}
\end{center}
\caption{\textbf{Experimentally measured noise powers.} The measured noise
powers of the output modes with a vacuum and a phase squeezed state as
inputs are plotted here. Black lines are SNL, blue and red lines are output
noise power without and with cluster state as ancillary state, yellow lines
are ideal output noise power. Squeezing degree of the squeezing gate is $-$%
12 dB. Measurement frequency is 2 MHz, the spectrum analyzer resolution
bandwidth is 30 kHz, and the video bandwidth is 100 Hz.}
\end{figure}

Fig. 5 shows the measured noise variances of the quadratures ($\hat{x}_{\mu
} $, $\hat{p}_{\mu }$, $\hat{x}_{\nu }$, and $\hat{p}_{\nu }$) of the output
modes ($\mu $ and $\nu $) with a vacuum mode $\alpha $ and a phase squeezed
state $\beta $\ with squeezing of $-4$ dB as two inputs of the gate
sequence, where $-$12 dB squeezing is implemented on input $\alpha $. In the
ideal case with the cluster state of infinite squeezing (yellow lines), the
noise powers of $\hat{x}_{\mu }$\ and $\hat{x}_{\nu }$ are 12 and 4 dB above
the SNL (black lines), which correspond to the anti-squeezing noises
resulting from the squeezing gate (12 dB) and the input phase squeezed state
(4 dB), respectively. The noise powers of $\hat{p}_{\mu }$\ and $\hat{p}%
_{\nu }$\ are 4.1 and 12.1 dB above the SNL due to the effect of CZ gate
[see equation (17)]. The blue and red lines stand for the output noises
without and with the cluster state as ancillary state, respectively. The
blue lines are obtained by using a coherent light of identical intensity to
replace each of cluster submodes. In this case, the measured values of $\hat{%
x}_{\mu }$, $\hat{p}_{\mu }$, $\hat{x}_{\nu }$, and $\hat{p}_{\nu }$ are $%
12.75\pm 0.07$, $9.05\pm 0.09$, $7.76\pm 0.08$ and $13.07\pm 0.09$ dB above
the SNL, respectively. The noise variances of $\hat{x}_{\mu }$, $\hat{p}%
_{\mu }$, $\hat{x}_{\nu }$, and $\hat{p}_{\nu }$ measured with the existence
of the cluster state (red lines) are $12.34\pm 0.11$, $7.60\pm 0.13$, $%
6.84\pm 0.12$ and $12.55\pm 0.13$ dB above the SNL, respectively, all of
which are lower than the corresponding values without using the cluster
resource.

\begin{figure*}[tbp]
\par
\begin{center}
\includegraphics[width=150mm]{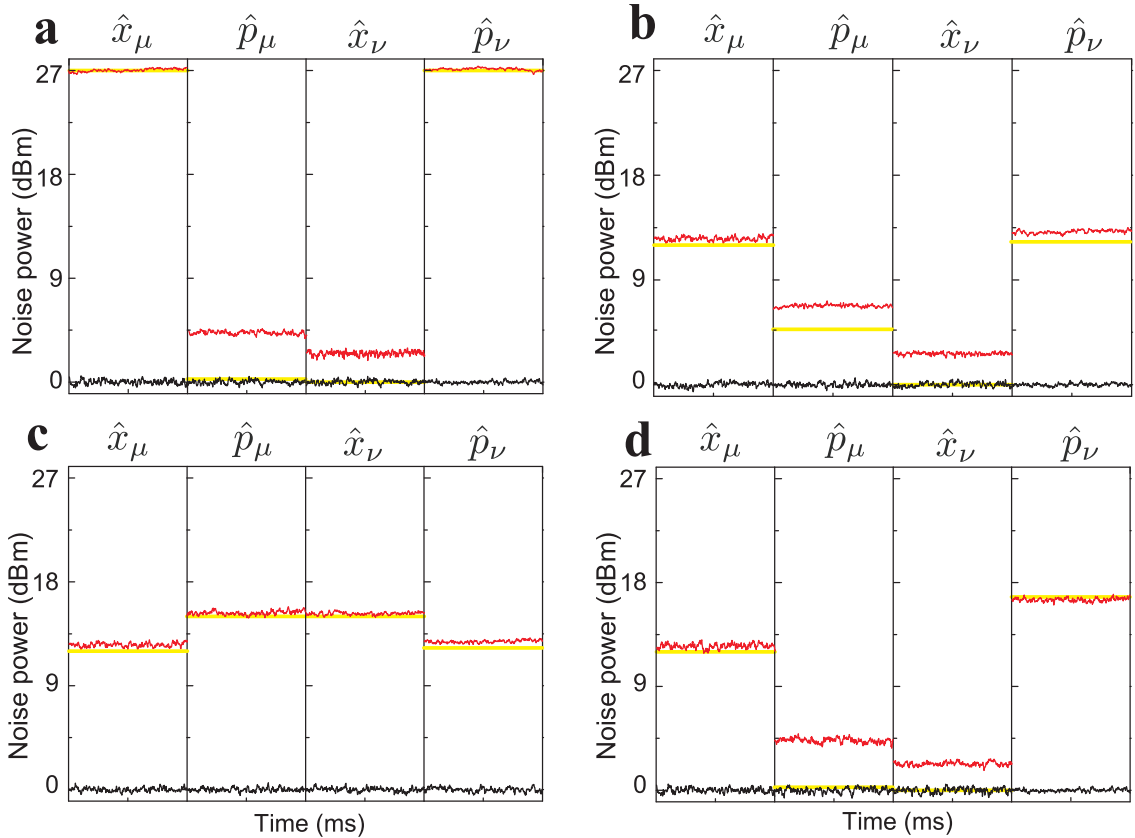} 
\end{center}
\caption{\textbf{Experimentally measured noise powers of the output modes
with coherent inputs.} \textbf{a-d}: $\hat{x}_{\protect\alpha }$, $\hat{p}_{%
\protect\alpha }$, $\hat{x}_{\protect\beta }$, and $\hat{p}_{\protect\beta }$%
-coherent state as input, respectively. Black lines are SNL, red lines are
output noise power with cluster state as ancillary state, yellow lines are
ideal output noise power. Squeezing degree of the squeezing gate is $-$12
dB. Measurement frequency is 2 MHz. The spectrum analyzer resolution
bandwidth is 30 kHz, and the video bandwidth is 100 Hz.}
\end{figure*}

In order to further verify the general input-output relation of the gate
sequence, we employ a coherent state with a 15 dB modulation signal as input
state (Fig. 6a-d). Fig. 6a shows the noise powers of quadratures of the
output modes $\mu $ and $\nu $ when input modes $\alpha $ and $\beta $ are a
coherent state with nonzero amplitude of 15 dB ($\hat{x}_{\alpha }$%
-coherent) and a vacuum state, respectively. The measured noise variance of $%
\hat{x}_{\mu }$ (red line) is $27.01\pm 0.13$ dB above SNL (black lines)
that is because 12 dB anti-squeezing noise resulting from squeezing gate is
added on the 15 dB input amplitude of $\hat{x}_{\alpha }$. In the ideal case
(yellow lines), the noise variance of $\hat{p}_{\mu }$ is a little higher
than SNL since $\hat{x}_{\nu }$ is added on the squeezed noise of $-$12 dB,
and the noise variance of $\hat{x}_{\nu }$ is at the level of SNL. The
measured noise powers of $\hat{p}_{\mu }$ and $\hat{x}_{\nu }$ is about $%
4.43\pm 0.16$ and $2.68\pm 0.18$ dB above the SNL because of the effect of
excess noises coming from the imperfect entanglement of the cluster state.
The measured noise variance of $\hat{p}_{\nu }$ is $27.02\pm 0.11$ dB above
the SNL because the amplitude on $\hat{x}_{\mu }$ is added to $\hat{p}_{\nu
} $, which satisfy the input-output relation of the CZ logic gate in
equation (17). Fig. 6b shows the noise powers of output modes when a
coherent state with a modulation signal of 15 dB on $\hat{p}_{\alpha }$ ($%
\hat{p}_{\alpha }$-coherent) and a vacuum state are used for the input
states $\alpha $\ and $\beta $, respectively. The measured noise power (red
lines) of $\hat{x}_{\mu }$ is $12.34\pm 0.17$ dB above the corresponding SNL
(black line) because of the effect of anti-squeezing noise resulting from
the squeezing gate. The noise power of $\hat{p}_{\mu }$ and $\hat{x}_{\nu }$
(red lines) are $6.72\pm 0.12$ dB and $2.68\pm 0.12$ dB above the
corresponding SNL, respectively. The noise power of $\hat{p}_{\nu }$ is $%
12.68\pm 0.14$ dB above the SNL because the noise of $\hat{x}_{\mu }$ is
added on $\hat{p}_{\nu }$ [see equation (17)]. Figs. 6c-d are the noise
powers of output modes when the input is the coherent state with the
modulation signal of 15 dB on $\hat{x}_{\beta }$, and $\hat{p}_{\beta }$ ($%
\hat{x}_{\beta }$-coherent and $\hat{p}_{\beta }$-coherent), respectively.
It is obvious that the measured noise powers of output modes satisfy the
input-output relation of the gate sequence in equation (17).

\begin{figure}[tbp]
\centerline{
\includegraphics[width=\columnwidth]{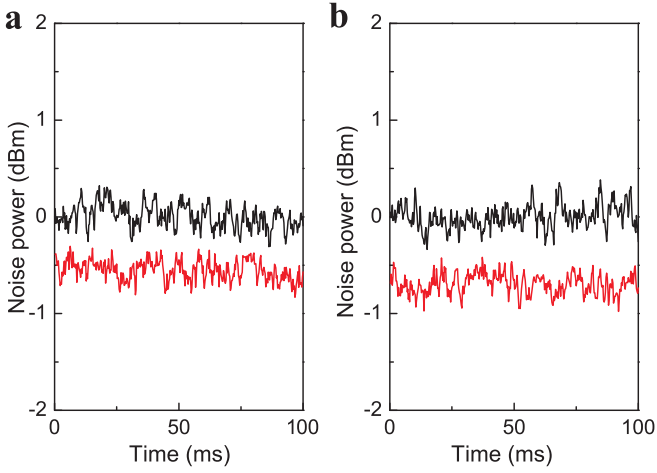}
}
\caption{\textbf{Measured quantum correlation variances.} This figure show
the measured quantum correlation variances of the output modes of the gate
sequence with a vacuum state and a phase squeezed state as inputs. \textbf{a}
and \textbf{b} are noise power of $\left\langle \Delta ^{2}(g\hat{p}_{%
\protect\mu }-\hat{x}_{\protect\nu })\right\rangle $ and $\left\langle
\Delta ^{2}(g\hat{p}_{\protect\nu }-\hat{x}_{\protect\mu })\right\rangle $,
respectively. Black lines and red lines are SNL and quantum correlation
noise, respectively. Squeezing degree of the squeezing gate is $-$12 dB.
Measurement frequency is 2 MHz, the spectrum analyzer resolution bandwidth
is 30 kHz, and the video bandwidth is 100 Hz.}
\end{figure}

Fig. 7 shows the measured correlation noise variances of the output modes
with a vacuum mode ($\alpha $) and a $-4$ dB phase squeezed mode ($\beta $)
as the inputs of the sequence, where $-12$ dB squeezing is implemented on
input $\alpha $. The measured noise variance (red lines) of $\left\langle
\Delta ^{2}(g\hat{p}_{\mu }-\hat{x}_{\nu })\right\rangle $ (a) and $%
\left\langle \Delta ^{2}(g\hat{p}_{\nu }-\hat{x}_{\mu })\right\rangle $ (b)
are $0.53\pm 0.11$ and $0.65\pm 0.11$ dB below the corresponding SNL (black
lines), respectively. The entanglement is quantified by
\begin{equation}
\left\langle \Delta ^{2}(g\hat{p}_{\mu }-\hat{x}_{\nu })\right\rangle
+\left\langle \Delta ^{2}(g\hat{p}_{\nu }-\hat{x}_{\mu })\right\rangle
=0.856\pm 0.022.
\end{equation}%
For our experimental system the calculated optimal gain factor $g=0.98$. The
total correlation variances in the left side of equation (18) is smaller
than $g$ and thus satisfies the inseparability criteria, which confirms the
quantum entanglement between the two output modes ($\mu $ and $\nu $) from
the gate sequence.

We also use fidelity $F=\left\{ \text{Tr}[(\sqrt{\hat{\rho}_{1}}\hat{\rho}%
_{2}\sqrt{\hat{\rho}_{1}})^{1/2}]\right\} ^{2}$, which denotes the overlap
between the experimentally obtained output state $\hat{\rho}_{2}$\ and the
ideal output sate $\hat{\rho}_{1}$, to quantify the performance of the gate
sequence. The fidelity for two Gaussian states $\hat{\rho}_{1}$\ and $\hat{%
\rho}_{2}$\ with covariance matrices $\mathbf{A}_{j}$\ and mean amplitudes $%
\mathbf{\alpha }_{j}\equiv (\alpha _{jx},\alpha _{jp})$\ ($j=1,2$)\ is
expressed by \cite{Nha2005,Scutaru1998} 
\begin{equation}
F=\frac{2}{\sqrt{\Delta +\sigma }-\sqrt{\sigma }}\exp [-\mathbf{\epsilon }%
^{T}(\mathbf{A}_{1}+\mathbf{A}_{2})^{-1}\mathbf{\epsilon }],
\end{equation}%
\ where $\Delta =\det (\mathbf{A}_{1}+\mathbf{A}_{2}),$\ $\sigma =(\det
\mathbf{A}_{1}-1)(\det \mathbf{A}_{2}-1),$\ $\mathbf{\epsilon }=\mathbf{%
\alpha }_{2}-\mathbf{\alpha }_{1},$ $\mathbf{A}_{1}$ and $\mathbf{A}_{2}$
for the ideal ($\hat{\rho}_{1}$) and experimental ($\hat{\rho}_{2}$) output
states, respectively. The covariance matrices $\mathbf{A}_{j}$\ ($j=1,2$)
for target mode are given by

\begin{eqnarray}
\mathbf{A}_{\mu 1} &=&4\left[
\begin{array}{cc}
\left\langle \Delta ^{2}\hat{x}_{\mu }\right\rangle _{1} & 0 \\
0 & \left\langle \Delta ^{2}\hat{p}_{\mu }\right\rangle _{1}%
\end{array}%
\right] ,\qquad \\
\mathbf{A}_{\mu 2} &=&4\left[
\begin{array}{cc}
\left\langle \Delta ^{2}\hat{x}_{\mu }\right\rangle _{2} & 0 \\
0 & \left\langle \Delta ^{2}\hat{p}_{\mu }\right\rangle _{2}%
\end{array}%
\right] .
\end{eqnarray}%
The coefficient \textquotedblleft 4\textquotedblright\ comes from the
normalization of SNL. Since the noise of a vacuum state is defined as 1/4 in
the text, while in the fidelity formula the vacuum noise is normalized to
\textquotedblleft 1\textquotedblright , so a coefficient \textquotedblleft
4\textquotedblright\ appears in the expressions of covariance matrices.
Similarly, we can write out the covariance matrices for the control mode

\begin{eqnarray}
\mathbf{A}_{\nu 1} &=&4\left[
\begin{array}{cc}
\left\langle \Delta ^{2}\hat{x}_{\nu }\right\rangle _{1} & 0 \\
0 & \left\langle \Delta ^{2}\hat{p}_{\nu }\right\rangle _{1}%
\end{array}%
\right] ,\qquad \\
\mathbf{A}_{\nu 2} &=&4\left[
\begin{array}{cc}
\left\langle \Delta ^{2}\hat{x}_{\nu }\right\rangle _{2} & 0 \\
0 & \left\langle \Delta ^{2}\hat{p}_{\nu }\right\rangle _{2}%
\end{array}%
\right] .
\end{eqnarray}%
In case of infinite squeezing, both fidelities for the target and control
states, $F_{\mu }$ and $F_{\nu }$, equal to 1, which can be calculated from
equation (6)\ with $r\rightarrow \infty $.

\begin{figure*}[tbp]
\setlength{\belowcaptionskip}{-3pt}
\centerline{
\includegraphics[width=150mm]{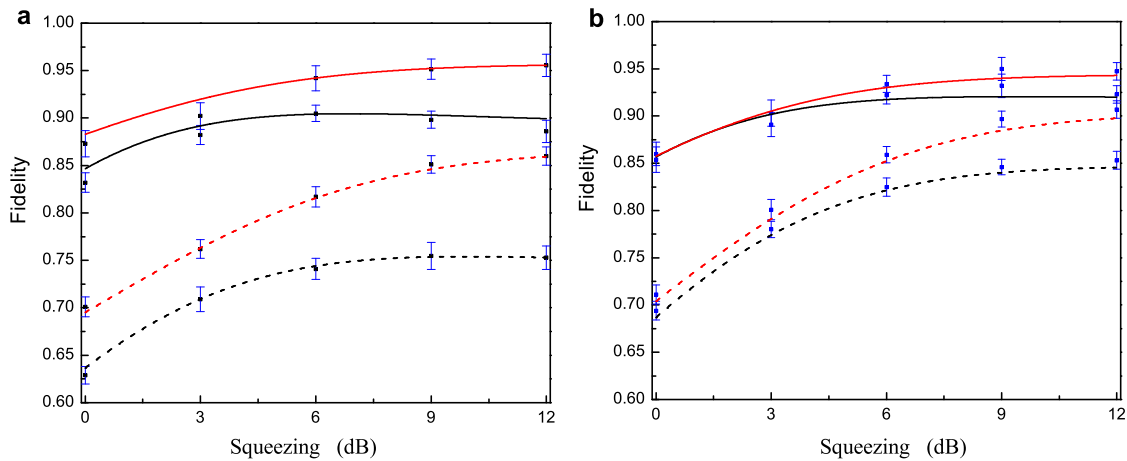}
}
\caption{\textbf{Dependence of fidelity on squeezing of the squeezing gate.}
\textbf{a} and \textbf{b} are the fidelity for the input mode $\protect\beta
$ as a vacuum state and a $-$4 dB phase squeezed state, respectively. The
solid and dashed lines correspond to fidelity with and without cluster
resource, respectively. Black and red lines present the fidelity of output
modes $\protect\mu $ and $\protect\nu $, respectively. Error bars represent $%
\pm$ one standard deviation and are obtained based on the statistics of the
measured noise variances.}
\end{figure*}

Fig. 8 is the fidelities of the output modes $\mu $ and $\nu $ as the
function of squeezing degree of the squeezing gate based on the experimental
data for two different input $\beta $ state (a: a vacuum state, b: a $-4$ dB
phase squeezed state). In Fig. 8, black and red lines correspond to fidelity
of output modes $\mu $ and $\nu $, respectively. Dashed lines are obtained
at the case without the use of cluster resource (using the coherent states
in the same intensity to substitute the cluster states in Fig. 1) and solid
lines are completed under the case using cluster resource state. Obviously,
when the cluster state is applied, the fidelity of the output states is
higher than that obtained at the case using the coherent state, which is
usually named as the classical limit in quantum optics. The experimentally
measured data (see table 1) are marked in Fig. 8a and 8b which are in
reasonable agreement with the theoretical calculation.

\section*{Discussion}

\begin{figure}[tbp]
\setlength{\belowcaptionskip}{-3pt}
\centerline{
\includegraphics[width=\columnwidth]{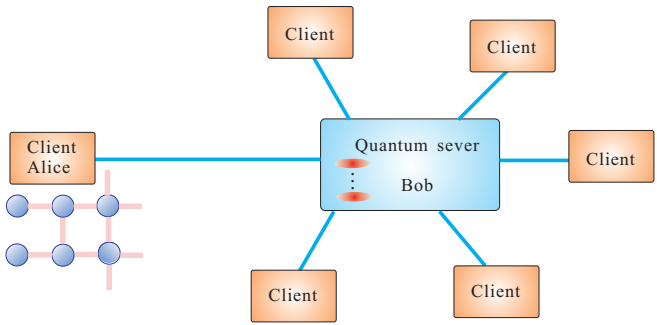}
}
\caption{\textbf{The schematic of reverse blind CVQC.} The quantum server
prepares squeezed states and sends them to anyone of clients in demand.
Client prepares CV cluster state with linear optics and then implements
quantum computation via measurements and feedforwards on the prepared
cluster state.}
\end{figure}

We demonstrated a gate sequence in one-way QC fashion by applying a six-mode
CV cluster state as quantum resource. The quantum feature of the gate
sequence is verified quantitatively by both the inseparability criterion of
two-mode entanglement and the fidelities of output states. The entanglement
degree of two output modes depends on two cascaded gates, simultaneously,
which exhibits the sequence character of the system. Although in our
experiment only two gates are linked together, the scheme can be easily
extended to construct any large QC gate sequence with a number of gates.

Today, quantum computers have become a physical reality and are continuing
to be developed. One-way QC based on quantum teleportation \cite%
{vanLoockJOSA2007,Gu2009} is able to implement secure information processing
and accomplish the unbreakable quantum coding \cite{Broadbent,Barz}. On the
other hand, the large cluster states can be prepared only by linearly
optical systems if appropriate squeezed states are available. Thus, one-way
quantum computers consisting of this type of gate sequences can be operated
in a reverse blind CVQC model to realize the secure QC network (Fig. 9), in
which only a server owns the ability of preparing quantum states (such as
squeezed states) and all remote clients ask the server to send them
necessary squeezed states through a quantum channel \cite{Mor}. Then,
clients produce the cluster states using linearly optical transformation and
perform arbitrary CV one-way Gaussian QC by means of classical measurements
and feedforwards on the prepared cluster state at their stations. In this
way, the server and any eavesdroppers never know what clients want to do,
thus the security of the blind QC is guaranteed by no-signaling principle
\cite{Pop}. The presented gate sequence for one-way Gaussian CVQC is an
essentially experimental exploring toward developing universal QC and
practical quantum networks.

\section*{Methods}

\textbf{Experimental details.} The three $\hat{x}$-squeezed and three $\hat{p%
}$-squeezed states are produced by three NOPAs. These NOPAs are pumped by a
common laser source, which is a continuous wave intracavity
frequency-doubled and frequency-stabilized Nd:YAP/LBO(Nd-doped YAlO$_{3}$
perorskite/lithium triborate) \cite{WangIEEE2010}. The output fundamental
wave at 1080 nm wavelength from the laser is used for the injected signals
of NOPAs and the local oscillators of the homodyne detectors (HDs). The
output second-harmonic wave at 540 nm wavelength serves as the pump field of
the four NOPAs, in which a pair of signal and idler modes with the identical
frequency at 1080 nm and the orthogonal polarizations are generated through
an intracavity frequency-down-conversion process \cite{Wang20102}. Each of
NOPAs consists of an $\alpha $-cut type-II KTP crystal and a concave mirror
\cite{Wang20102}. The front face of the KTP was coated to be used for the
input coupler and the concave mirror serves as the output coupler of the
squeezed states, which is mounted on a piezo-electric transducer for locking
actively the cavity length of NOPA on resonance with the injected signal at $%
1080$ nm. The transmissions of the input coupler at 540 nm and 1080 nm are $%
99.8\%$ and $0.04\%$, respectively. The transmissions of the output coupler
at 540 nm and 1080 nm are $0.5\%$ and $5.2\%$, respectively. The finesses of
the NOPA for 540 nm and 1080 nm are $3$ and $117$, respectively. In our
experiment, all NOPAs are operated at the parametric deamplification
situation, i.e. the phase difference between the pump field and the injected
signal is $(2n+1)\pi $ ($n$ is an integer). Under this condition, the
coupled modes at +45$^{\circ }$ and -45$^{\circ }$ polarization directions
are the quadrature-amplitude and the quadrature-phase squeezed states,
respectively \cite{Su2007,Yun2000}.

Three NOPAs are locked individually by means of Pound-Drever-Hall (PDH)
method with a phase modulation of 56 MHz on 1080 nm laser beam \cite{Pound}.
In the experiment, the relative phase (2n+1)$\pi $ locking is completed with
a lock-in amplifier, where a signal around 15 kHz is modulated on the pump
light by the piezo-electric transducer (PZT) mounted on a reflection mirror
which is placed in the optical path of the pump laser and then the error
signal is fed back to the other PZT which is mounted on a mirror placed in
the optical path of the injected beam. In the beam-splitter network used to
prepare six-mode cluster states, the relative phase between two incident
beams on T1 and T4 is phase-locked to zero, and that on T2, T3 and T5 is
phase-locked to $\pi /2$. The zero phase difference (T1 and T4) between two
interfered beams on a beam-splitter is locked by a lock-in amplifier. The $%
\pi /2$ phase difference (T2, T3 and T5) is locked by DC locking technique,
where the photocurrent signal of the interference fringe is fed back to the
PZT mounted on a mirror which is placed before the beam-splitter. In the
homodyne detection system, zero phase difference for the measurement of
quadrature-amplitude is locked by PDH technique with a phase modulation of
10.9 MHz on local oscillator beam. The $\pi /2$ phase for the measurement of
quadrature-phase is locked with DC locking technique too.

The transmission efficiency of an optical beam from NOPA to a homodyne
detector is around 96\%. The quantum efficiency of a photodiode
(FD500W-1064, Fermionics) used in the homodyne detection system is 95\%. The
interference efficiency on a beam-splitter is about 99\%. The phase
fluctuation of the phase locking system is about 2-3$%
{{}^\circ}%
$\textbf{.}

\section*{Acknowledgments}

This research was supported by the National Basic Research Program of China
(Grant No. 2010CB923103), NSFC (Grant Nos. 11174188, 61121064) and OIT.

\end{document}